\documentclass[apjl]{emulateapj}
\usepackage{apjfonts}

\received{August 18, 2008}
\accepted{January 16, 2009}

\slugcomment{Accepted by the Astrophysical Journal Letters} 
\shorttitle{Crystallization Physics from WDs}
\shortauthors{Winget et al.}

\begin{document}
\title{The Physics of Crystallization from Globular Cluster White
  Dwarf Stars in NGC 6397}

\author{ D.\ E.\ Winget\altaffilmark{1,2}, S.\ O.\
  Kepler\altaffilmark{2}, Fab\'{\i}ola Campos\altaffilmark{2}, M.\ H.\
  Montgomery\altaffilmark{1,3}, Leo Girardi\altaffilmark{4}, 
P. Bergeron\altaffilmark{5},
Kurtis Williams\altaffilmark{1,6}}

\altaffiltext{1}{Department of Astronomy, University of Texas at
  Austin, Austin, TX, USA; dew@astro.as.utexas.edu}
\altaffiltext{2}{Instituto de F\'{\i}sica, Universidade Federal do Rio
  Grande do Sul, Porto Alegre, RS - Brasil} 
\altaffiltext{3}{Delaware Asteroseismic Research Center, Mt.\ Cuba
  Observatory, Greenville, DE, USA}
\altaffiltext{4}{INAF -
  Padova Astronomical Observatory} 
\altaffiltext{5}{D\'epartement de Physique, Universit\'e de
  Montr\'eal, C. P. 6128, Succ. Centre-Ville, Montr\'eal, Qu\'ebec H3C
  3J7, Canada}
\altaffiltext{6}{NSF Astronomy \& Astrophysics Postdoctoral Fellow}

\begin{abstract}
  
We explore the physics of crystallization in the deep interiors of white dwarf stars using the color-magnitude diagram and luminosity function constructed from proper motion cleaned Hubble Space Telescope photometry of the globular cluster NGC~6397.  We demonstrate that the data are consistent with the theory of crystallization of the ions in the interior of white dwarf stars and provide the first empirical evidence that the phase transition is first order: latent heat is released in the process of crystallization as predicted by \citet{VanHorn68}. We outline how this data can be used to observationally constrain the value of $\Gamma\equiv E_\mathrm{Coulomb}/E_\mathrm{thermal}$ near the onset of crystallization, the central carbon/oxygen abundance, and the importance of phase separation.

\end{abstract}

\section{Star Formation History and Physics from the White Dwarf Stars}

White dwarf stars are the inevitable progeny of nearly all ($\simeq
97$\%) stars \citep[e.g.,][ hereafter FBB]{Fontaine01}. Their
distribution can be used to extract two things: age of the stellar
population and cooling physics of the white dwarf (WD) stars.  The two are
interrelated, but qualitatively different.  Extracting the age and
history of star formation has become known as WD
cosmochronology.  An excellent review emphasizing this connection and
the attendant uncertainties is given by FBB.

The techniques of WD cosmochronology have been successfully
applied to the disk by a number of investigators \citep[e.g.,][and
references therein]{Winget87,Wood92,Hansen03} and are being
continuously refined.  They have also been applied to a variety of
open clusters and calibrated against main sequence turnoff and related
methods \citep[e.g.,][and references therein]{Kalirai07,DeGennaro08}.
The Hubble Space Telescope (HST) photometry obtained by Richer and
Hansen and their collaborators \citep{Hansen02,Hansen07,Richer08} has
yielded a new harvest of information for WD populations.
They have used the Advanced Camera on the HST to reach the terminus of
the WD cooling sequence, giving us a qualitatively different
tool for analyzing the WD population.  \citet{Hansen07} used
Monte Carlo techniques in conjunction with their cooling models to
determine the age of NGC~6397 from the WD stars, attempting
to account for uncertainties in the basic physical parameters of the
WD stars to determine an age for the cluster; using goodness
of fit criteria, they arrive at an age for the cluster, based on 
WD cooling, of $11.47 \pm 0.47$~Gyr.


Finding the signature of the key physical properties of the WD stars
in the disk luminosity function (hereinafter LF) has proven more
difficult than getting an age constraint.  This is because the disk
population contains stars formed at different times and from different
main sequence progenitors.  This is greatly simplified in a cluster
sample and, most of all, in an old globular cluster.  In this paper we
focus on the HST photometry of NGC~6397 and the distribution of WD
stars in the color magnitude diagram.  We report evidence for a
``bump'' in the LF due to the release of the latent heat of
crystallization and we show how this can be refined to yield more
accurate measures of these processes.

\section{Anchoring the White Dwarf Evolutionary Sequences in the Color-Magnitude Plane}

We fit main sequence, pre-WD, and WD evolutionary
models simultaneously. The main sequence and pre-WD models we
used for this work were computed with the Padova stellar evolution
code \citep{Marigo08}. We used a variety of metallicities to determine
the best fit to the main sequence and WD models.

Our WD evolutionary models have updated constitutive physics
\citep[see e.g.,][]{Kim08a}.  We place the new generation of WD
evolutionary models of DA and DB WD stars in the
observed $F_{814W}$ vs.\ $F_{606W}$ color-magnitude diagram
(hereinafter CMD) using P. Bergeron's model atmosphere
grids\footnote{http://www.astro.umontreal.ca/$\sim$bergeron/CoolingModels/}
\citep[for a detailed description
see][]{Bergeron95,Holberg06,Holberg08} along with an analytical
correction to the \citet{Kowalski07} results for the effect of
Ly~$\alpha$ far red-wing absorption.  This correction is small and
will be discussed in a forthcoming paper.

\begin{figure}[t]
  \centering{
    \hspace*{-1em}
\includegraphics[width=1.05\columnwidth,angle=0]{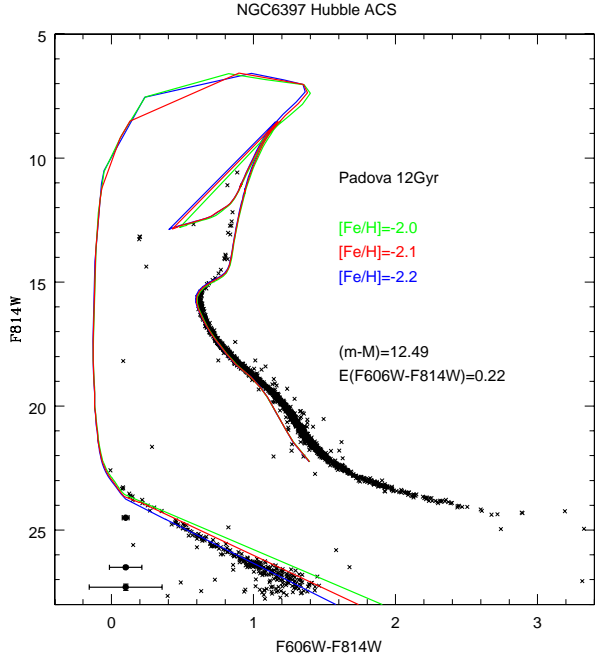}
}
    \vspace*{-1em}
\caption{ The best fit to the proper-motion screened HST data on
  NGC~6397 using the Padova stellar evolution code \citep{Marigo08}
  and the Bergeron (2006) atmospheres.  This fit gives ${\rm
    [Fe/H]}=-2.2$, $(m-M)=12.49$, and $E(F606W-F814W)=0.22$. This
  anchors our WD sequences in the CMD.  }
\label{cmd}
\end{figure}

Our best fit of the CMD in the natural ACS color system gives a
metallicity of $Z=0.00012\pm 0.00001$, $E(606W-814W)=0.22\pm 0.02$,
and $(m-M)=12.49\pm 0.05$ (Figure~1) and a main sequence turnoff age
of $12^{+0.5}_{-1.0}$~Gyr.  The age is consistent with the values found
by \citet{Richer08} and \citet{Hansen07}, even though they used the
Dartmouth Evolutionary Sequence (DES). The metallicity is a factor of
two lower than \citet{Richer08} but is in agreement with the
independent direct spectroscopic determinations based on VLT data
\citep{Korn07}. The values of these parameters fix the WD cooling
tracks in the color-magnitude diagram and eliminate the freedom to
slide the tracks as has been done in other works \citep[e.g.,
][]{Hansen07}. This constrains the best-fit total WD mass throughout
the CMD. It is clear from Figure~2 that the tracks fit the bulk of the
sample well.

The data (see Figure~2) are taken from the proper motion selected
sample of \citet{Richer08}. This is a more homogeneous sample than
that of \citet{Hansen07} but is smaller because of the reduced area
and magnitude limits of the proper-motion data.  Four features stand
out in Figure~2: there is a gap in the distribution near $F814W =
24.5$ that may be statistically significant, there is a noticeable
concentration of stars near $F814W = 26.5$, there is a terminus at
approximately $F814W = 27.6$ and a noticeable turn to the blue before
the terminus.  These last two features were noted in \citet{Hansen07}.
In this paper we focus on what we can learn from the concentration, or
clump, of stars near 26.5, providing a physical explanation.

\section{Physics With the CMD and Luminosity Function}

The CMD diagram constrains the mechanical and thermal properties of
the WD stars \citep{Richer08}.  Once the evolutionary tracks
have been anchored by the main sequence and WD sequence
simultaneously as described above we can move on to exploring the
physics contained in the CMD. \citet{Hansen07} point out that the
location of the terminus provides a simple lower-limit to the age of
the cluster from the WD cooling times. For our models, this
WD cooling limit is reached at about $10.5$~Gyr for pure
carbon core models. This age limit is consistent with the values
quoted in \citet{Hansen07} and \citet{Richer08}.  The position of the
tracks is insensitive to processes affecting only the age; to examine
these we must look to the LF, the number of stars
observed as a function of magnitude.

\begin{figure}[t]
\includegraphics[angle=-90,width=1.02\columnwidth]{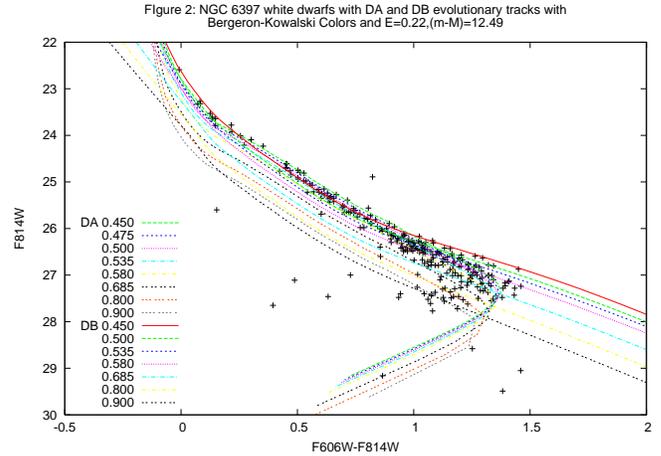}
\caption{ NGC~6397 WDs with DA and DB 
  evolutionary sequences using Bergeron (2006) atmospheres adjusted
  with an analytical fit to the \citet{Kowalski07} changes for
  Ly-$\alpha$ red-wing opacity. 
  }
\label{cmdmix}
\end{figure}

For the parameters described above, it is evident from Figure~2 that
models with masses in the range of $0.500-0.535 M_{\sun}$ best fit the
region near the center of the clump. This increases slightly for
decreased values of $m-M$ and is also a function of the reddening.  We
emphasize again that simultaneously fitting the main sequence and
white dwarf sequence provides tight constraints on both the distance
modulus and the reddening.

\begin{figure}[b]
  \centering{
\includegraphics[width=1.0\columnwidth]{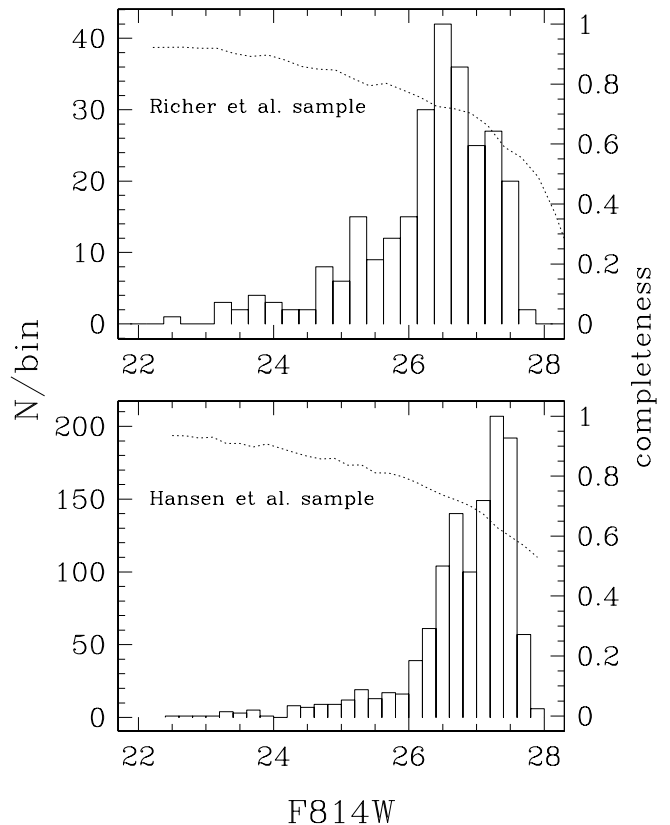}
}
\caption{top panel: The observed WDLF (histogram)
  and completeness relation (dotted line) of the \citet{Richer08}
  sample for NGC~6397. lower panel: the same for the \citet{Hansen07}
  sample. For both samples, we note that the completeness changes
  fairly slowly over the region of the observed rapid fall-off of
  stars and while remaining above 50\%.  }
  \label{comp}
\end{figure}

In Figure~3 we show the LF of both the \citet{Hansen07} and
\citet{Richer08} samples.  The peak in both LFs near $F814W=26.5$
suggests that evolution slows through this region.  In the sample of
\citet{Hansen07} the LF continued to rise well past this peak, so it
was not the maximum of the distribution; it completely dominates the
distribution after the application of the proper motion selection
\citep{Richer08}.  The completeness estimates of both samples have
been carefully considered by the respective authors (Figure~3, dotted
lines) and the relatively slow variation of the completeness near this
peak implies it is not the result of incompleteness. We therefore seek
a physical explanation of this peak in the context of a physical
process that occurs near this point in the models.

Two processes occur in the dominant DA models near this point:
crystallization and the convective coupling (e.g., FBB).
Crystallization, through the release of latent heat, slows down
evolution and produces a bump in the LF. 
Convection, when it reaches down to the degeneracy boundary,
decreases the insulation of the nondegenerate envelope and temporarily
increases the total temperature gradient; this serves to slow down the
evolution, briefly, then causes it to speed up again. This produces a
broad feature in the WD cooling curve that will have a
signature in the LF.


\subsection{The $\,\Gamma$ of Crystallization}

Crystallization in the dense Coulomb plasma of WD interiors
was theoretically predicted independently by \citet{Kirzhnits60},
\citet{Abrikosov60}, and \citet{Salpeter61} to occur when the ratio of
the Coulomb energy to the thermal energy of the ions (the ratio
``$\Gamma$'') is large. For a one-component plasma (OCP), there is
universal agreement among different theoretical approaches that
crystallization occurs when $\Gamma \simeq 175$ \citep[e.g.,
][]{Slattery82,Stringfellow90,Potekhin00,Horowitz07}. There is a
similar consensus, based largely on a density-functional approach,
that this value of $\Gamma$ also holds for a binary carbon and oxygen
mixture.  Such a mixture is likely relevant to WD interiors.
Recently, \citet{Horowitz07} used a massive molecular dynamics
computation to explore crystallization in a dense Coulomb plasma.
They found $\Gamma =175$ for an OCP, while for a specific mixture of
elements they found $\Gamma \simeq 237$. As we show, such a difference
is potentially measurable from the observations of WD stars
in globular clusters or older open clusters.

\subsection{Luminosity functions with and without crystallization}

\citet{Hansen07} demonstrated in their analysis that the entire
observed sequence represents a very narrow range of WD masses,
including magnitudes well below $F814W=26.5$. It is therefore
reasonable for purposes of this initial exploration to adopt a
fiducial mass. On the basis of the model tracks in the CMD shown in
Figure~2, we choose the model that passes nearest the color of the red
edge of the clump of stars corresponding to the peak in the LF; this
model has a mass of $0.5 M_{\sun}$. For the layer masses, we assume
$M_{\rm H}/M_{\star} = 10^{-4}$ and $M_{\rm He}/M_{\star} = 10^{-2}$
for the DA sequences, and $M_{\rm He}/M_{\star} = 10^{-2}$ for the DB
sequence. We adopt a carbon core model including the effects of
crystallization for this sequence.

Assuming a constant star formation rate, the theoretical LF is
proportional to the ``cooling function'' of an evolutionary model
sequence. This function is given by the derivative $(dt/dm)$, where
$m$ is the F814W magnitude of a given model and $t$ is its age. Since
we will be comparing directly with the data, we also multiply the
theoretical LF by the completeness correction given explicitly in
Table~4 of \citet{Richer08} and shown in the top panel of our
Figure~3. Finally, we normalize the resulting curves by minimizing
their root-mean-square residuals in the neighborhood of the peak,
between an F814W of 25.1 and 27.5.

\begin{figure}[t]
\vspace*{-1em}
  \centering{
\includegraphics[width=1.03\columnwidth]{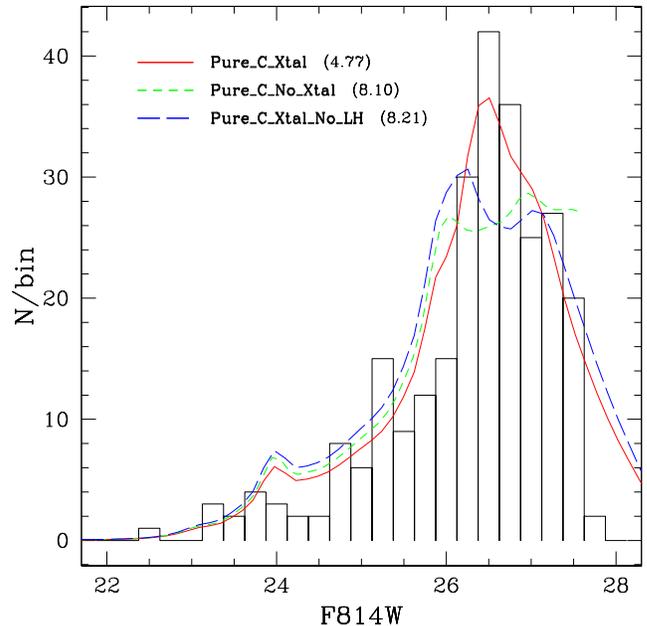}
}
\vspace*{-1em}
\caption{ The observed WDLF of NGC~6397 \citep[histogram]{Richer08}
  with LFs from theoretical evolutionary sequences of $0.5 M_{\sun}$
  DA models with pure carbon cores (lines): crystallization with
  (``Pure\_C\_Xtal'') and without (``Pure\_C\_Xtal\_No\_LH'') the
  release of latent heat, and excluding the physics of crystallization
  altogether (``Pure\_C\_No\_Xtal'').  The normalization of the
  theoretical curves is chosen to minimize the RMS residuals in the
  neighborhood of the peak, between the magnitudes of 25.1 and 27.7,
  the faintest value calculated for the no crystallization case. The
  value of the average residual for each curve is listed in the
  legend, e.g., it is 4.77 for the ``Pure\_C\_Xtal'' case.}
  \label{pureC}
\end{figure}

In Figure~4 we show the LF of our fiducial sequence
(``Pure\_C\_Xtal''), that of a sequence with crystallization
artificially suppressed (``Pure\_C\_No\_Xtal''), and that of one
including crystallization but artificially excluding the latent heat
of crystallization (``Pure\_C\_Xtal\_No\_LH''); all are plotted over a
histogram of the observed LF. The no crystallization and no latent
heat sequences show evidence of a bump due to convective coupling
around $F814W \sim 26$, but do not continue to rise through the
observed maximum; this is clearly inconsistent with the data, and no
adjustment of mass or internal composition can bring them into good
agreement. In terms of $\chi^2$, for average observational errors of
$\sim 5.5$ stars/bin in the neighborhood of the peak, we have
$\chi^2=0.75$ for the crystallizing sequence and $\chi^2 \sim 2.2$ for
the no crystallization and no latent heat sequences, a nearly
three-fold increase in $\chi^2$.  Thus, the sequence with
crystallization provides a much better match to the data.

\subsection{Constraining crystallization, phase separation, and core
  composition}

In the years since \citet{VanHorn68} the realization that the cores of
normal mass WD stars should consist of a mixture of carbon and oxygen
implied that crystallization may also release energy resulting from
phase separation of the carbon and oxygen
\citep{Stevenson77,Barrat88,Segretain93,Segretain94,Isern00}. This
occurs because when a carbon/oxygen mixture crystallizes the oxygen
content of the solid should be enhanced. Since WDs crystallize from
the center outwards, this leads to a net transport of oxygen inward
and carbon outward, and because oxygen is slightly heavier than carbon
this differentiation releases gravitational energy.

\begin{figure}[t]
\vspace*{-1em}
  \centering{
\includegraphics[width=1.03\columnwidth]{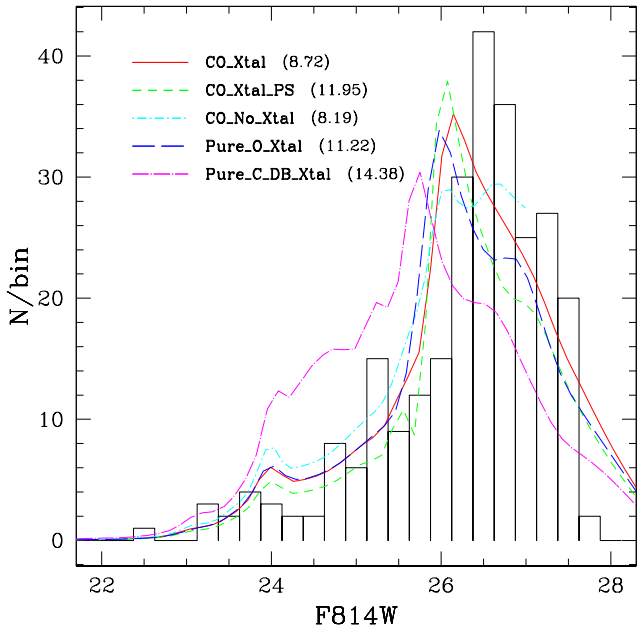}
}
\vspace*{-1em}
\caption{ The same as Figure~\ref{pureC} but for DA model sequences
  with uniform, 50:50 carbon/oxygen cores: crystallization only
  (``CO\_Xtal''), crystallization \emph{and} phase separation
  (``CO\_Xtal\_PS''), and no crystallization (``CO\_No\_Xtal'').  In
  addition, we show a pure oxygen DA sequence (``pure\_O\_Xtal'') and
  a pure carbon DB sequence (``pure\_C\_DB\_Xtal'').  All models have
  $0.5M_{\sun}$.}
\label{NGC 6397 white dwarf luminosity function}
\end{figure}

We have included this energy in our models as described in
\citet{Montgomery99}. For these computations we have assumed the
carbon and oxygen abundances are equal throughout the core.  This
underestimates the oxygen abundance compared to that predicted by
standard stellar evolution calculations \citep[e.g.,][]{Salaris97},
but the remaining uncertainty in the C$(\alpha,\gamma)$O reaction rate
\citep{Metcalfe02,Assuncao06} lead to a degree of uncertainty in the
C/O ratio and profile. In Figure~5 we show several LFs:
crystallization only (``CO\_Xtal''), crystallization with phase
separation (``CO\_Xtal\_PS''), no crystallization (``CO\_No\_Xtal''),
pure oxygen core with crystallization (``Pure\_O\_Xtal''), and a pure
carbon DB sequence (``Pure\_C\_DB\_Xtal''). As is readily apparent,
all of these sequences have a peak which is too bright by at least
0.5~mag in $F814W$.

These results have several interesting possible interpretations.
First, the results seem to suggest that the oxygen
content of these stars is relatively small or zero, since it is the
higher crystallization temperature of oxygen which shifts the peak in
the LF to smaller magnitudes.  The only way to accommodate more oxygen
would be to have lower mass models, in conflict with the distance modulus
\citep{Hansen07}. The colors (e.g., Figure~2) also make it difficult
to appeal to lower masses with higher oxygen abundances.
Additionally, for plausible IFMRs the main sequence lifetime for
single stars becomes more problematic with lower WD masses even with
the IFMR dependency on metallicity of \citet{Meng08}. Thus the
constraint on the interior oxygen abundance becomes stronger. Taken at
face value, these results indicate that the carbon-to-oxygen ratio is
much greater than 1, and we will be able to make a more quantitative
statement from our future more complete Bayesian statistical analysis
(in preparation).

Second, as shown in Figure~4, the data are consistent and well-fit by
carbon core models \emph{with} crystallization and the release of
latent heat, but not by models without. This confirms the prediction
of \citet{VanHorn68} that crystallization is a first order phase
transition and releases the latent heat of crystallization. Were it
not so, crystallization would leave no sharp peak at this magnitude in
the observed LF.  This impacts our understanding of solid-state
physics at extremely high density: it is the first empirical
confirmation of the release of latent heat during crystallization --
an important theory that has a large impact on WD ages, as has been
pointed out by \citet{VanHorn68} and many authors in the intervening
years.

Third, it is possible a priori that a significant fraction of the
observed WDs may be DBs.  The mismatch of observed LF and the DB
sequence in Figure~5 essentially eliminates pure He atmospheres as a
significant component of the sample \citep[as shown by][]{Hansen07},
but not models that become mixed (H/He) as they cool; we explore this
possibility in a forthcoming paper.

Fourth, the ``best-fit'' fiducial sequence in Figure~4 begins
crystallizing near the value of $\Gamma_{\rm xtal} \equiv
E_\mathrm{Coulomb}/E_\mathrm{thermal} \sim 170$. If the actual value
is higher (lower) then crystallization will occur at lower (higher)
luminosities. Higher values allow fits with larger amounts of oxygen
in their cores. In addition, \citet{Potekhin00} show that even for a
pure composition theoretical uncertainties in the polarization of the
electron Fermi gas and quantum effects in the liquid and solid phase
can alter the value of $\Gamma_{\rm xtal}$. In future analyses, an
accurate determination of the mass, distance, and reddening will lead
to an accurate determination of $\Gamma_{\rm xtal}$ and the core
composition.


Finally, we note that the central density and temperature associated
with a particular value of $F814W$ through the model atmospheres is
sensitive only to the mass-radius relationship set by the degenerate
electron pressure support.  This is very insensitive to the $C/O$
relative abundances --- these produce differences of $\delta F814W <
0.05$.  This implies that the value of $\Gamma$ in the center, at the
peak of the LF for example, is sensitive only to the interior
composition.  Therefore we conclude that the onset of crystallization
is determined by the particular mixture and the value of $\Gamma$ for
that mixture.  Comparison of the theoretical models and the data
promise to provide important measures of the onset and development of
crystallization.

\section{Discussion, Summary, and Futures of Exploring White Dwarf
  Physics with CMDs}

Although we are not focused on uncertainties, it is reasonable to
examine how changing the distance modulus might affect the results.
Put another way, how much does the distance have to change to
reproduce the peak of the observed LF with oxygen crystallization
rather than carbon?  The answer is contained in Figure~5. Here we see
that to make the location of the peak of the LF consistent with oxygen
crystallization we have to lower the distance modulus by a little more
than $0.5$ magnitudes --- this possibility is excluded by the main
sequence fitting \citep{Richer08}.


We have shown that simultaneously fitting the main sequence and the
WDs in a cluster gives the best possible constraint on
distance, metallicity and reddening corrections.  Physically realistic
atmosphere calculations then allow us to place evolutionary tracks in
the CMD.  The number distribution of stars contains important
information on the internal physics of the WD stars.  This
allows us to explore the physics of crystallization. We present
evidence that the data is most consistent with a first order phase
transition, releasing latent heat during crystallization, as proposed
by \citet{VanHorn68}. The current data places constraints on the onset
of crystallization, the central carbon/oxygen abundance, and the
composition of the envelope at the degeneracy boundary. We will
improve these constraints with a more complete Bayesian statistical
analysis in the near future.  This work also points to the importance
of forthcoming data on additional clusters as well as increasing the
sample of stars through more HST fields on this cluster.  This work
also underscores the essential nature of more proper motion data to
get the most information out of these kinds of studies.

Pulsations may also allow an asteroseismological determination of the
crystallized mass fraction for massive pulsators, as shown by
\citet{Metcalfe04} for the DAV BPM~37093, although this claim has been
challenged by \citet{Brassard05}. While certainly important, we note
that asteroseismological analyses do not probe the latent heat of
crystallization; this quantity is accessible only through the WDLF, as
demonstrated in this paper. To this end, we eagerly anticipate the
forthcoming HST observations of this cluster.  These will provide a
proper-motion screened sample over a larger area of the cluster and to
fainter magnitudes, providing an exacting test of the ideas put forth
in this paper.

\acknowledgments The authors wish to thank A. Zabot for help computing
the isochrones, Ted von Hippel, and Hugh van Horn for useful
discussions, and Brad Hansen and referees for suggesting improvements.  D.E.W., S.O.K., and F.C. are fellows of CNPq - Brazil.
M.H.M. and K.A.W. are grateful for the financial support of the
National Science Foundation under awards AST-0507639 and AST-0602288,
respectively, and M.H.M. acknowledges the support of the Delaware
Asteroseismic Research Center. This work was supported in part by the
NSERC Canada and by the Fund FQRNT (Qu\'ebec). P. Bergeron is a
Cottrell Scholar of Research Corporation.


\clearpage

\end{document}